\documentclass[preprint]{aastex}
\newcommand{\bdv}[1]{\mbox{\boldmath$#1$}}
\def\bpi{{\bdv{\pi}}}

\title{WFIRST Planet Masses from Microlens Parallax}

\author{
J.C.~Yee\altaffilmark{1}
}
\altaffiltext{1}{Department of Astronomy, Ohio State University, 140
W. 18th Ave., Columbus, OH 43210, USA;
jyee@astronomy.ohio-state.edu}

\begin{document}

\begin{abstract}
  I present a method using only a few ground-based observations of
  magnified microlensing events to routinely measure the parallaxes of
  WFIRST events if WFIRST is in an L2 orbit. This could be achieved
  for all events with $A_{\rm max}>30$ using target-of-opportunity
  observations of select WFIRST events, or with a complementary,
  ground-based survey of the WFIRST field, which can push beyond this
  magnification limit. When combined with a measurement of the angular
  size of the Einstein ring, which is almost always measured in
  planetary events, these parallax measurements will
  routinely give measurements of the lens masses and hence, the
  absolute masses of the planets. They can also lead to mass
  measurements for dark, isolated objects such as brown dwarfs,
  free-floating planets, and stellar remnants if the size of the
  Einstein ring is measured.

\end{abstract}

\section{Introduction}

The microlensing portion of the WFIRST mission will complete the
census of planets by finding large populations of planets beyond the
snow line with masses as small as Mars \citep{Green12}. If the
masses of the planets and their hosts are measured, this will permit a
direct comparison to planet formation theories. However, the primary
observables in microlensing events are the mass ratio and projected
separation (scaled to the Einstein ring) between the planet and its
host star. A measurement of the lens mass is necessary to transform
these to physical quantities.

If the lens is bright enough, WFIRST will be able to estimate its mass
based on a measurement of the lens flux. However, there will be many
cases for which the lens light will be too faint to be measured. Such
cases will most likely be lenses at the bottom of the stellar mass
function, but could also include brown dwarfs, free-floating planets,
or stellar remnants. For these events, with only an upper limit on the
lens flux, the conclusions that can be drawn about the nature of the
planet are limited. In addition, the WFIRST measurement of the lens
flux will be a measurement of the total flux of the lens system,
including any companions to the lens, which may or may not participate
in the lensing event. Typically, companions within 10 AU will produce
a measurable microlens perturbation and companions at more than 5 mas
($\sim 40$ AU) will be identifiable from a shift in the centroid
relative to the lensing event. However, companions at intermediate
separations are not easily identified. Hence, the WFIRST mass
estimate will necessarily be an upper limit for any given system.

Fortunately, the lens masses can be measured if microlens parallax and
finite source effects are observed. Microlens parallax is a vector
quantity whose magnitude is the ratio of Earth's orbit to the
size of the Einstein ring projected onto the observer plane,
$\tilde{r}_{\rm E}$:
\begin{equation}
\label{eqn:pimag}
\pi_{\rm E} \equiv \frac{\rm AU}{\tilde{r}_{\rm E}}.
\end{equation}
If $\pi_{\rm E}$ is measured, the mass of the lens, $M$, can be obtained
with a measurement of the angular size of the Einstein ring, $\theta_{\rm E}$:
\begin{equation}
M=\left(\frac{\theta_{\rm E}}{\pi_{\rm E}}\right)\left(\frac{c^2{\rm AU}}{4G}\right).
\end{equation}
For any event in which the size of the source is resolved in time,
i.e., it passes over a caustic or near a cusp, $\theta_{\rm E}$ is
measurable. Such finite source effects are almost always measured in
events with planets because detection of a planetary companion to the
lens almost always requires a caustic interaction. Hence, if the
microlens parallax can be measured, the planet masses are
known. Finite source effects can also be measured in any event for
which the source crosses the position of the lens.

In this Letter, I discuss a means to routinely measure the lens masses
using microlens parallax if WFIRST is in an L2 orbit\footnote{See
  \citet{Gould13} for a discussion of WFIRST parallax measurements for
  a geocentric orbit.}. Because the
WFIRST light curve will be measured so precisely, the orbital parallax
effect will be routinely detected at high significance, effectively
giving one extremely well-measured component of the parallax
\citep{Gould13}. I show that only a few ground-based observations of
each event are needed to complement the WFIRST observations and yield
a complete parallax measurement for a large fraction of events.

\section{Measuring $\pi_{\rm E,\perp}$}

\subsection{Simplified Case \label{sec:approx}}

\begin{figure}
\includegraphics[width=\textwidth]{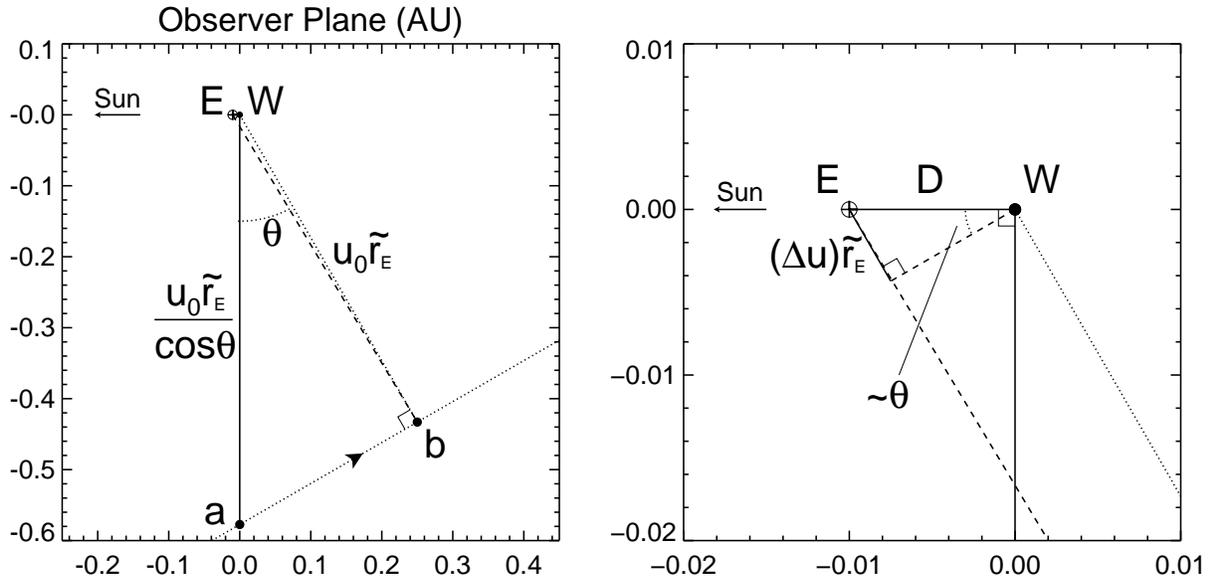}
\caption{Left panel: basic geometry of a microlensing event projected
  onto the observer plane. Right panel: expanded view around the
  projected positions of Earth and WFIRST(`E' and `W',
  respectively). The x-axis is parallel to the projection of the
  Sun-Earth-WFIRST line. The dotted line ab shows the lens
  trajectory.  The value of $\pi_{\rm E, \perp}({\rm AU})^{-1} =
  \sin\theta/\tilde{r}_{\rm E}=\Delta u/D$ can be derived from the
  observables $\Delta u$ and $D$, while $\pi_{\rm E,
    \|}=\cos\theta/\tilde{r}_{\rm E}$ can be measured by WFIRST
  alone. \label{fig:geom}}
\end{figure}

The microlens parallax vector can be written
\begin{equation}
\label{eqn:pidef}
  \bpi_{\rm E} = (\pi_{\rm E, \|}, \pi_{\rm E, \perp}) = (\pi_{\rm E}\cos\theta, \pi_{\rm E}\sin\theta),
\end{equation}
where $\theta$ is the angle between the lens trajectory and the
projection of the Sun-Earth line on the plane of the sky, measured
counter-clockwise. Because WFIRST will be in orbit about the Sun,
there will be a measurable asymmetry in the light curve due to the
orbital parallax effect \citep{Gould92a, GouldMB94}. This gives strong
constraints on the component of the parallax parallel to the projected
position of the Sun relative to the event, $\pi_{\rm E,\|}$, but
usually only very weak constraints on the other component, $\pi_{\rm
  E, \perp}$ \citep[e.g., ][]{Gould13}.

I will show that if WFIRST is at L2, as few as two observations of a
microlensing event from Earth can be used to measure $\pi_{\rm E,
  \perp}$, leading to a measurement of $\pi_{\rm E}$. I begin with a
simplified case to illustrate the problem. In the following section, I
will present the full derivation and expression for $\pi_{\rm E,
  \perp}$ and show that it reduces to what is derived here.

Consider the projection of the microlensing event onto the observer
plane (Fig. \ref{fig:geom}). For the purposes of illustration, I
assume that the size of Einstein ring projected onto this plane is
$\tilde{r}_{\rm E} = 10$ AU, $u_0=0.05$, and that the observations are
taken close to the equinox (the anticipated midpoint of WFIRST
observations) so that the projection of the Earth-WFIRST line onto the
sky, $D$, is equal to the true separation, i.e., $D=0.01$ AU. The
exact values for these quantities are irrelevant to the derivation; I
will discuss their practical implications below. Finally, note that
WFIRST at L2 puts it in line with Earth and Sun, so the projection of
the Earth-WFIRST line on the sky is parallel to the projection of the
Sun's position.

The WFIRST light curve will be extremely well measured, giving
$\pi_{\rm E,\|}$ and the basic microlens parameters: the time of the
peak, the source-lens impact parameter scaled to the Einstein ring,
and the Einstein crossing time ($t_0$, $u_0$, and $t_{\rm E}$,
respectively). Hence, the value of $u_0 \tilde{r}_{\rm
  E}(\cos\theta)^{-1}$ is also known. This fixes point `a' on the lens
trajectory projected onto the observer plane. Assume the event is
observed from Earth when it is at the peak as seen from WFIRST
(i.e., when the lens is at point `b'). Then, the fractional difference
in the magnification is 
\begin{equation}
  \frac{\Delta A}{A} = \frac{A-A_{\oplus}}{A} \simeq \frac{\Delta u}{u_0}
\end{equation}
where $A$ is the magnification as seen from WFIRST, $A_{\oplus}$ is
the magnification as seen from Earth, and I assume that the
magnification is given by $A\simeq u^{-1}_0$ (which applies in the
limit $u_0\ll1$) and that the difference between the impact parameter
as seen from Earth and from WFIRST is $\Delta u \ll u_0$. As
illustrated in Figure \ref{fig:geom}, in the regime where
$u_0\tilde{r}_{\rm E} \gg D$, $(\Delta u)\tilde{r}_{\rm E}\simeq
D\sin\theta$ meaning that with some manipulation $\pi_{\rm E,\perp}$
can be written:
\begin{equation}
\label{eqn:pieperp}
  \pi_{\rm E, \perp} = u_0\left(\frac{\Delta A}{A}\right)\left(\frac{\rm AU}{D}\right).
\end{equation}
Note that all the variables in the right-hand side of the equation are
known or measurable.

From the geometry in Figure \ref{fig:geom}, there is one
degeneracy in Equation (\ref{eqn:pieperp}). It is possible to change
the sign of $u_0$, i.e., reflect the figure over the x-axis, which
changes the sign of $\pi_{\rm E, \perp}$. This leads to a degeneracy
in the direction of $\bpi_{\rm E}$, but not in its magnitude, which is
the relevant quantity for calculating masses. Out of the eight
possible configurations one might consider as potentially degenerate
with the geometry shown, only the $u_0 \rightarrow -u_0$ degeneracy
described here is permitted by the observables.

\subsection{Exact Expression for $\pi_{\rm E,\perp}$ 
\label{sec:app}}

\begin{figure}
\includegraphics[width=\textwidth]{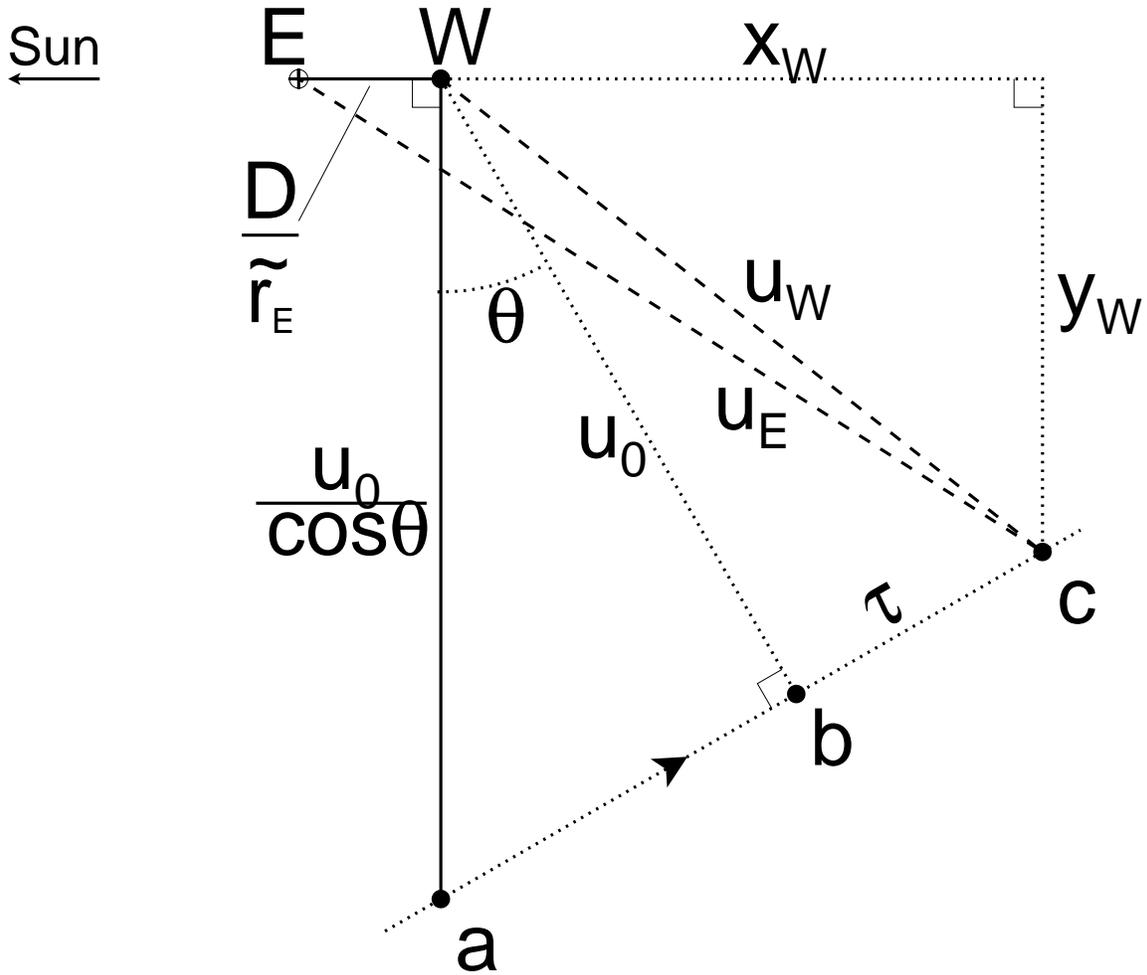}
\caption{Generalized geometry of a microlensing event projected onto
  the Observer plane. This figure is analogous to Fig. \ref{fig:geom}
  except that all quantities are scaled to the size of
  the Einstein ring, $\tilde{r}_{\rm E}$. The line abc indicates the
  trajectory of the lens. The projected positions of Earth and
  WFIRST are labeled `E' and `W', respectively.\label{fig:figA}}
\end{figure}

I now derive a general expression for $\pi_{\rm E, \perp}$ that
applies for an Earth-based measurement of the magnification at any
time. In practice, this is the expression that will be used to
calculate $\pi_{\rm E,\perp}$ from the observables.

Figure \ref{fig:figA} shows the generalized geometry with all
quantities scaled to $\tilde{r}_{\rm E}$. Consider an observation
from Earth is taken at time $t$, i.e., when the lens is at point `c.'
The lens position is given by $u_0$ and $\tau = (t-t_0)t_{\rm E}^{-1}$. The
measured magnification is related to the separation of the lens, $u$,
by
\begin{equation}
A(u)=\frac{u^2+2}{u\sqrt{u^2+4}},
\end{equation}
where $u$ is measured as a fraction of the Einstein ring. Thus, from
the measured magnifications as seen from Earth and WFIRST, the lens
separation from their projected positions is known, $u_{\rm E}$ and
$u_{\rm W}$, respectively. I can write
\begin{equation}
\label{eqn:triangle}
u_{\rm E}^2 = (x_{\rm W}+D/\tilde{r}_{\rm E})^2 + y_{\rm W}^2,
\end{equation}
where $x_{\rm W}$ and $y_{\rm W}$ are the projections of $u_{\rm W}$ onto the x-
and y-axes. Equation (\ref{eqn:triangle}) can be rewritten
as
\begin{equation}
  u_{\rm E}^2 - u_{\rm W}^2 = \frac{2D}{\tilde{r}_{\rm E}}(u_0\sin\theta+\tau\cos\theta)+\frac{D^2}{\tilde{r}_{\rm E}^2}.
\end{equation}
Recognizing that $(\pi_{\rm E, \|}, \pi_{\rm E, \perp}) = (\cos\theta,
\sin\theta){\rm (AU)}\tilde{r}_{\rm E}^{-1}$ (Eqn. \ref{eqn:pimag}
and \ref{eqn:pidef}), I evaluate
\begin{equation}
\pi_{\rm E,\perp} = \left[\Delta u (2u_{\rm W} +\Delta u)-\frac{D^2}{\tilde{r}_{\rm E}^2}\right]\left(\frac{\rm AU}{2Du_0}\right)-\frac{\tau\pi_{\rm E,\|}}{u_0},
\end{equation}
where $\Delta u \equiv u_{\rm E} - u_{\rm W}$. Since
$\tilde{r}_{\rm E}^{-2}=(\pi_{\rm E,\perp}^2+\pi_{\rm E, \|}^2) ({\rm
  AU})^{-2}$, this can be rewritten as a quadratic equation for
$\pi_{\rm E, \perp}$ with the solutions:
\begin{equation}
\label{eqn:exact}
  \pi_{\rm E,\perp,\pm} = u_0\left(\frac{\rm AU}{D}\right)\left(-1\pm\sqrt{1-\frac{1}{u_0^2}\left[\left(\frac{D}{\rm AU}\right)^2\pi_{\rm E, \|}^2+ 2\left(\frac{D}{\rm AU}\right)\tau\pi_{\rm E, \|} - \Delta u(2u_{\rm W}+\Delta u) \right]}\right).
\end{equation}
Although there are formally two solutions for $\pi_{\rm E, \perp}$,
these can readily be distinguished. The solution $\pi_{\rm E, \perp,
  -}$ corresponds to the case in which the lens passes between the
projected positions of Earth and WFIRST. This scenario is expected to
be very rare, but it can be definitively excluded with additional
observations of the event from Earth.

If $\pi_{\rm E}$ is large, then the full expression must be evaluated.
However, the present Letter is primarily focused on cases for which
$\pi_{\rm E}$ is small because those are the cases in which the WFIRST
light curve will constrain only one component of the parallax well. In
that case, Equation (\ref{eqn:exact}) is well represented by the first
term in the Taylor expansion:
\begin{equation}
  \pi_{\rm E,\perp,+} = \frac{1}{2u_0^2}\left[\Delta u(2u_{\rm W}+\Delta u)- 2\left(\frac{D}{\rm AU}\right)\tau\pi_{\rm E, \|}-\left(\frac{D}{\rm AU}\right)^2\pi_{\rm E, \|}^2 \right].
\end{equation}
Furthermore, the last term can generally be ignored because it is
second order. Finally, if the event is observed at peak
($\tau\rightarrow 0$ and $u_{\rm W} \rightarrow u_0$) and we assume that
$|\Delta u|\ll |u_0|$, 
\begin{equation}
  \pi_{\rm E,\perp} \rightarrow \Delta u \left(\frac{\rm AU}{D}\right),
\end{equation}
which is equivalent to Equation (\ref{eqn:pieperp}).

\subsection{Constraints on $\pi_{\rm E, \perp}$}

The uncertainties from the WFIRST light curve are negligible compared
to the uncertainties from the ground-based photometry, so the largest
uncertainty in $\pi_{\rm E, \perp}$ comes from the measurement of
$(\Delta A) A^{-1}$. The actual observables from Earth are the
magnified flux, $f_{\rm mag, \oplus}$, and flux of the event at the
baseline, $f_{\rm base, \oplus}$, such that
\begin{equation}
  A_{\oplus} = \frac{f_{\rm mag,\oplus}-f_{\rm base, \oplus}}{f_{\rm
      S, \oplus}}+1,
\end{equation}
where $f_{\rm S, \oplus}$ is the source flux as seen from Earth. To
solve for $A_{\oplus}$, the unknown source flux must be estimated by
calibrating the ground-based photometry to the WFIRST photometry using
comparison stars. This situation is equivalent to the problem of
measuring $\Delta u$. \citet{Gould95}, \citet{BoutreuxGould96}, and
\citet{GaudiGould97} showed that $\Delta u$ is poorly constrained
because $u_0$ is correlated with $t_{\rm E}$ and $f_{\rm B}$, where
$f_{\rm B} = f_{\rm base} - f_{\rm S}$ is the blended (non-varying) component of
the flux. However, most of the information about $f_{\rm B}$ comes
from the wings of the event \citep{Yee12}, which will not be well
measured from the ground because the sky background is so high. Hence,
flux calibration  is necessary to find $f_{\rm S}$, constrain $f_{\rm
  B}$, and improve the precision of the measurement of
$u_0$, or equivalently $A_{\oplus}$.

If only one or two observations are taken of the
magnified event, the flux calibration ultimately sets the
limit on the precision of $\pi_{\rm E, \perp}$. Based on previous
experience \citep[e.g.,][]{Yee12}, the uncertainty in this calibration
is limited by systematics to a precision of about 1\%. This sets the
fundamental noise floor on the measurement of $(\Delta A) A^{-1}$.
Given that by definition, $|(\Delta u)\tilde{r}_{\rm E}| \leq D$, the
limit in the flux precision means that a $3\sigma$ measurement of
$\pi_{\rm E, \perp}$ is possible for
\begin{equation}
\label{eqn:phot}
u_0 \leq 0.03 \left(\frac{D}{0.01 {\rm AU}}\right)\left(\frac{\tilde{r}_{\rm E}}{10 {\rm AU}}\right)^{-1}\left(\frac{\sigma_{\Delta A/A}}{0.01}\right)^{-1}
\end{equation}
where I again make the assumption that $u_0\ll1$, i.e., $A\simeq
u_0^{-1}$. 

In contrast, if there are many magnified points that can be seen above
the baseline from the ground, the peak of the event will be resolved
allowing a measurement of the effective timescale, $t_{\rm eff}=u_0
t_{\rm E}$. \citet{Yee12} showed that this quantity is invariant to
uncertainties in $f_{\rm B}$, so flux calibration is
unnecessary. Because the velocity offset between WFIRST at L2 and
Earth are quite small (0.3 km s$^{-1}$), $t_{\rm E}$ is approximately
the same for the WFIRST and ground-based light curves. Then, the
uncertainty in $\Delta u$ is:
\begin{equation}
\sigma_{\Delta u} \simeq \sigma_{u_{0,\oplus}} = u_{0,\oplus}\sqrt{\left(\frac{\sigma_{t_{\rm eff,\oplus}}}{t_{\rm eff, \oplus}}\right)^2+\left(\frac{\sigma_{t_{\rm E}}}{t_{\rm E}}\right)^2} ,
\end{equation}
where $u_{0,\oplus}$ is the impact parameter of the event as seen from
Earth and I assume that $u_{\rm 0, sat}$ is known essentially
perfectly. Hence, the method can be applied to events with $u_0>0.03$
by reducing the uncertainty in $t_{\rm eff, \oplus}$ using additional observations.

Finally, I note that even if $\pi_{\rm E, \perp}$ is less well
measured than $\pi_{\rm E, \|}$, this does
not mean that the value of $\pi_{\rm E}$ is not well measured. So long
as $\pi_{\rm E, \|} \gtrsim 3\sigma_{\pi_{\rm E, \perp}}$, the
constraints on $\pi_{\rm E}$ will be useful. This will be true for a
large fraction of cases, depending on how well the lens trajectory
aligns with the projection of the Sun-Earth-WFIRST line, which is
primarily a random effect.

\section{Discussion}

For pure satellite parallax measurements, the Earth-L2 baseline is not
ideal because it is only a small fraction of $\tilde{r}_{\rm E}$ (0.01
AU vs. $\sim 10$ AU), limiting the precision of the measurement that
can be made. However, here I take advantage of several consequences of
this special geometry that had not been previously considered for a
microlensing satellite at L2 \citep{GouldGaudiHan03, Han04} and the
precision of the WFIRST light curve, which will allow the $\pi_{\rm E,
  \|}$ component of the parallax to be measured extremely well. The
short baseline actually resolves the magnitude degeneracy described in
\citet{Gould94b} because it is extremely unlikely that the lens will
pass in between the two observatories, and if such a case occurred,
the parallax would be so large as to be easily measured from the
WFIRST light curve alone. In addition, L2 is in line with the
projected position of the Sun, which when combined with the
measurement of $\pi_{\rm E, \|}$ greatly simplifies the geometry
(Figures \ref{fig:geom} and \ref{fig:figA}). Finally, because L2 is
moving with Earth, the relative velocity offset between Earth and the
satellite can be neglected relative to stellar motions (0.3 km
s$^{-1}$ versus 300 km s$^{-1}$). This means that $t_{\rm E}$ is
essentially the same for the ground-based and space-based observations
and the degeneracies discussed in \citet{Gould99,Dong07} are avoided.

I have shown that for an event discovered by WFIRST at L2, a
measurement of the event magnification as seen from Earth yields a
measurement of the component of the parallax perpendicular to the
projection of Earth's orbit, $\pi_{\rm E, \perp}$. Although the
basic calculation was done assuming the event was observed from the
ground at the peak as seen from WFIRST, I showed that in principle the
Earth measurement can be made at any time. However, it is best to make
the measurement as close to the peak of the event as possible, since
the fractional difference in magnification will be largest at the
peak, allowing for the best measurement of the parallax. 

Measuring the magnification of the event as seen from Earth requires
at least two observations: one when it is magnified and one at
baseline. A third, magnified, observation would be beneficial in case
the source is passing over a caustic at the time of the observations
and for distinguishing between the two possible solutions for
$\pi_{\rm E, \perp}$ given in Equation (\ref{eqn:exact}). These
ground-based observations can be made either with
target-of-opportunity (ToO) observations of WFIRST events announced in
real time or with a simultaneous, ground-based survey of the WFIRST
microlensing fields.

If there are only a few observations, as would likely be the case with
ToO, $\pi_{\rm E,\perp}$ is measurable in all events with
$u_0\tilde{r}_{\rm E}<0.3 $AU, i.e., $u_0<0.03$ or peak magnification
$A_{\rm max}<30$. This limit is set by systematics in the flux
calibration between the ground-based data and the WFIRST data, which
experience shows is limited to a precision of 1\%. If only a few
events can be observed from the ground, it is best to focus on the
highest magnification events. In practice many of the WFIRST sources
will be quite faint, so while 1\% precision will be possible for the
brighter sources, the limits on $u_0$ are probably more stringent for
the majority of events. This leads to a preference for higher
magnification events. However, it is precisely these events for which
parallax measurements are most desirable because these events are the
most likely to have planetary signals \citep{Griest98}. Furthermore,
for point lens events, the higher the magnification, the more likely
it is that finite source effects will be observed, allowing mass
measurements for these lenses. These isolated objects could include
stellar remnants, isolated stellar mass black holes, brown dwarfs, or
the population of free-floating planets found by
\citet{Sumi11}. \citet{GouldYee13} also proposed a means to measure
the mass of free-floating planets using terrestrial parallax. However,
because the baseline for terrestrial parallax measurements is $\sim
R_{\oplus}$, parallaxes are only measurable for the closest
objects. Here, parallax measurements are possible for much more
distant lenses (and hence, a larger volume and larger number of
events) because of the larger Earth-WFIRST baseline. Hence, targeted
observations for measuring parallaxes should focus on the higher
magnification events.

A NIR, ground-based survey simultaneous with WFIRST has the distinct
advantage over ToO that it would not require real-time information. In
half of all events with planets, the planetary signal will occur after
peak, so the opportunity to measure the parallax with ToO will be
missed. In addition, a survey would take multiple points throughout
the light curve. At the very least, this will improve the precision of
the ground-based flux measurement, allowing the limit of 1\% precision
to be reached for more events. However, the peaks of the events may
also be resolved, leading to a measurement of $t_{\rm eff}$ and
allowing $\pi_{\rm E, \perp}$ to be measured for events with $A_{\rm
  max}<30$. Such a survey could be carried out with existing
facilities such as the UKIRT or VISTA telescopes or a purpose-built
facility. Either way, it would cost a fraction of the WFIRST mission
cost and yield substantial scientific benefit.

There is value in carrying out both a survey and a ToO program. The
brighter events would be covered by the survey, but ToO with a larger
telescope equipped with adaptive optics could reach fainter
events. Combining the two approaches would maximize the number of
events for which parallax observations can be made while minimizing
the cost.

Although the method for obtaining parallaxes described here is
observationally intensive, the potential scientific impact makes such
observations invaluable. The core accretion theory of planet formation
predicts that giant planets should be rare around M dwarfs
\citep{Laughlin04,Ida05}, which are typical microlensing hosts. Hence,
measuring the masses of the lens stars and planets allows a direct
comparison to the theory, which is otherwise very difficult if only
mass ratios are known. Furthermore, when the lens mass is measured,
solving:
\begin{equation}
  \tilde{r}_{\rm E} = \frac{\rm AU}{\pi_{\rm E}}=\sqrt{\frac{4GM}{c^2}\frac{D_{\rm L}D_{\rm S}}{D_{\rm S}-D_{\rm L}}},
\end{equation}
yields a measurement of its distance, $D_{\rm L}$ (where the source
distance, $D_{\rm S}$, is assumed to be in the bulge). Such
measurements would allow a comparison of the planet populations in the
bulge and disk. Given that the stars (and planets) in the bulge formed
in a dense region of rapid star formation, one might expect a dearth
of giant planets there \citep{Thompson12}. In addition, although
WFIRST will measure the lens system fluxes for many events, whether
the light comes from the lens itself or a stellar companion will be
unknown. Systematic measurements of the microlens parallax can be used
to measure the fraction of events for which the lens light is
contaminated by the presence of a companion not involved in the
microlensing event. Finally, if microlens parallax is measured for
many events, including ones with lens mass estimates from WFIRST, this
will allow the first systematic test of the parallax effect.

Although this Letter has been written from the perspective of
the WFIRST mission, it is broadly applicable to any microlensing
satellite at L2. Thus, if a microlensing survey is included in the
Euclid mission \citep[cf.][]{Penny13, BeaulieuTisserandBatista13}, it
would also benefit from complementary ground-based observations. In
fact, for Euclid a ground-based parallax campaign is even more
important for measuring the lens masses because its NIR resolution
will make it more difficult to accurately measure the lens system
fluxes.

\acknowledgments{I thank Matthew Penny, Ondrej Pejcha, and especially
  Andrew Gould for helpful conversations.}

\bibliographystyle{/home/morgan/jyee/tex/apj}

\begin{thebibliography}{21}
\expandafter\ifx\csname natexlab\endcsname\relax\def\natexlab#1{#1}\fi

\bibitem[{{Beaulieu} {et~al.}(2013){Beaulieu}, {Tisserand}, \&
  {Batista}}]{BeaulieuTisserandBatista13}
{Beaulieu}, J.~P., {Tisserand}, P., \& {Batista}, V. 2013, ArXiv:1303.6783B 

\bibitem[{{Boutreux} \& {Gould}(1996)}]{BoutreuxGould96}
{Boutreux}, T. \& {Gould}, A. 1996, \apj, 462, 705

\bibitem[{{Dong} {et~al.}(2007){Dong}, {Udalski}, {Gould}, {Reach}, {Christie},
  {Boden}, {Bennett}, {Fazio}, {Griest}, {Szyma{\'n}ski}, {Kubiak},
  {Soszy{\'n}ski}, {Pietrzy{\'n}ski}, {Szewczyk}, {Wyrzykowski}, {Ulaczyk},
  {Wieckowski}, {Paczy{\'n}ski}, {DePoy}, {Pogge}, {Preston}, {Thompson}, \&
  {Patten}}]{Dong07}
{Dong}, S., et al. 2007, \apj, 664, 862

\bibitem[{{Gaudi} \& {Gould}(1997)}]{GaudiGould97}
{Gaudi}, B.~S. \& {Gould}, A. 1997, \apj, 477, 152

\bibitem[{{Gould}(1992)}]{Gould92a}
{Gould}, A. 1992, \apj, 392, 442

\bibitem[{{Gould}(1994)}]{Gould94b}
---. 1994, \apjl, 421, L75

\bibitem[{{Gould}(1995)}]{Gould95}
---. 1995, \apjl, 441, L21

\bibitem[{{Gould}(1999)}]{Gould99}
---. 1999, \apj, 514, 869

\bibitem[{{Gould}(2013)}]{Gould13}
---. 2013, \apjl, 763, L35


\bibitem[{{Gould} {et~al.}(2003){Gould}, {Gaudi}, \& {Han}}]{GouldGaudiHan03}
{Gould}, A., {Gaudi}, B.~S., \& {Han}, C. 2003, \apjl, 591, L53

\bibitem[{{Gould} {et~al.}(1994){Gould}, {Miralda-Escude}, \&
  {Bahcall}}]{GouldMB94}
{Gould}, A., {Miralda-Escude}, J., \& {Bahcall}, J.~N. 1994, \apjl, 423, L105

\bibitem[{{Gould} \& {Yee}(2013)}]{GouldYee13}
{Gould}, A. \& {Yee}, J.~C. 2013, \apj, 764, 107

\bibitem[{{Green} {et~al.}(2012){Green}, {Schechter}, {Baltay}, {Bean},
  {Bennett}, {Brown}, {Conselice}, {Donahue}, {Fan}, {Gaudi}, {Hirata},
  {Kalirai}, {Lauer}, {Nichol}, {Padmanabhan}, {Perlmutter}, {Rauscher},
  {Rhodes}, {Roellig}, {Stern}, {Sumi}, {Tanner}, {Wang}, {Weinberg}, {Wright},
  {Gehrels}, {Sambruna}, {Traub}, {Anderson}, {Cook}, {Garnavich},
  {Hillenbrand}, {Ivezic}, {Kerins}, {Lunine}, {McDonald}, {Penny}, {Phillips},
  {Rieke}, {Riess}, {van der Marel}, {Barry}, {Cheng}, {Content}, {Cutri},
  {Goullioud}, {Grady}, {Helou}, {Jackson}, {Kruk}, {Melton}, {Peddie},
  {Rioux}, \& {Seiffert}}]{Green12}
{Green}, et al. 2012, arXiv:1208.4012

\bibitem[{{Griest} \& {Safizadeh}(1998)}]{Griest98}
{Griest}, K. \& {Safizadeh}, N. 1998, \apj, 500, 37

\bibitem[{{Han} {et~al.}(2004){Han}, {Chung}, {Kim}, {Park}, {Ryu}, {Kang}, \&
  {Lee}}]{Han04}
{Han}, C., {Chung}, S.-J., {Kim}, D., {Park}, B.-G., {Ryu}, Y.-H., {Kang}, S.,
  \& {Lee}, D.~W. 2004, \apj, 604, 372

\bibitem[{{Ida} \& {Lin}(2005)}]{Ida05}
{Ida}, S. \& {Lin}, D.~N.~C. 2005, \apj, 626, 1045

\bibitem[{{Laughlin} {et~al.}(2004){Laughlin}, {Bodenheimer}, \&
  {Adams}}]{Laughlin04}
{Laughlin}, G., {Bodenheimer}, P., \& {Adams}, F.~C. 2004, \apjl, 612, L73

\bibitem[{{Penny} {et~al.}(2012){Penny}, {Kerins}, {Rattenbury}, {Beaulieu},
  {Robin}, {Mao}, {Batista}, {Calchi Novati}, {Cassan}, {Fouque}, {McDonald},
  {Marquette}, {Tisserand}, \& {Zapatero Osorio}}]{Penny13}
{Penny}, M.~T., {Kerins}, E., {Rattenbury}, N., {Beaulieu}, J.-P., {Robin},
  A.~C., {Mao}, S., {Batista}, V., {Calchi Novati}, S., {Cassan}, A., {Fouque},
  P., {McDonald}, I., {Marquette}, J.~B., {Tisserand}, P., \& {Zapatero
  Osorio}, M.~R. 2012, ArXiv:1206.5296

\bibitem[{{Sumi} {et~al.}(2011){Sumi}, {Kamiya}, {Bennett}, {Bond}, {Abe},
  {Botzler}, {Fukui}, {Furusawa}, {Hearnshaw}, {Itow}, {Kilmartin}, {Korpela},
  {Lin}, {Ling}, {Masuda}, {Matsubara}, {Miyake}, {Motomura}, {Muraki},
  {Nagaya}, {Nakamura}, {Ohnishi}, {Okumura}, {Perrott}, {Rattenbury}, {Saito},
  {Sako}, {Sullivan}, {Sweatman}, {Tristram}, {Udalski}, {Szyma{\'n}ski},
  {Kubiak}, {Pietrzy{\'n}ski}, {Poleski}, {Soszy{\'n}ski}, {Wyrzykowski},
  {Ulaczyk}, \& {Microlensing Observations in Astrophysics (MOA)
  Collaboration}}]{Sumi11}
{Sumi}, T., et al. 2011, \nat, 473, 349

\bibitem[{{Thompson}(2013)}]{Thompson12}
{Thompson}, T.~A. 2013, MNRAS, in press

\bibitem[{{Yee} {et~al.}(2012){Yee}, {Shvartzvald}, {Gal-Yam}, {Bond},
  {Udalski}, {Koz{\l}owski}, {Han}, {Gould}, {Skowron}, {Suzuki}, {Abe},
  {Bennett}, {Botzler}, {Chote}, {Freeman}, {Fukui}, {Furusawa}, {Itow},
  {Kobara}, {Ling}, {Masuda}, {Matsubara}, {Miyake}, {Muraki}, {Ohmori},
  {Ohnishi}, {Rattenbury}, {Saito}, {Sullivan}, {Sumi}, {Suzuki}, {Sweatman},
  {Takino}, {Tristram}, {Wada}, {the MOA Collaboration}, {Szyma{\'n}ski},
  {Kubiak}, {Pietrzy{\'n}ski}, {Soszy{\'n}ski}, {Poleski}, {Ulaczyk},
  {Wyrzykowski}, {Pietrukowicz}, {the OGLE Collaboration}, {Allen}, {Almeida},
  {Batista}, {Bos}, {Christie}, {DePoy}, {Dong}, {Drummond}, {Finkelman},
  {Gaudi}, {Gorbikov}, {Henderson}, {Higgins}, {Jablonski}, {Kaspi}, {Manulis},
  {Maoz}, {McCormick}, {McGregor}, {Monard}, {Moorhouse}, {Mu{\~n}oz},
  {Natusch}, {Ngan}, {Ofek}, {Pogge}, {Santallo}, {Tan}, {Thornley}, {Shin},
  {Choi}, {Park}, {Lee}, {Koo}, \& {the {$\mu$}FUN Collaboration}}]{Yee12}
{Yee}, J.~C., et al. 2012, \apj, 755, 102

\end{thebibliography}



\end{document}